\documentclass[12pt,preprint]{aastex}
\usepackage[numberedappendix]{emulateapj5}
\usepackage{epsf}
\usepackage{apjfonts}
\usepackage{graphicx}
\usepackage{timesfonts}

\newcommand{\bx}{{\bf x}}
\makeatletter
\newenvironment{inlinetable}{%
\def\@captype{table}%
\noindent\begin{minipage}{0.999\linewidth}\begin{center}\footnotesize}
{\end{center}\end{minipage}\smallskip}

\newenvironment{inlinefigure}{%
\def\@captype{figure}%
\noindent\begin{minipage}{0.999\linewidth}\begin{center}}
{\end{center}\end{minipage}\smallskip}
\makeatother

\begin{document}   

\submitted{\rm\it NAOJ-Th-Ap2001, No.18}

\title{Galaxy number counts in the Hubble Deep Field as a strong constraint on
a hierarchical galaxy formation model}

\author{Masahiro Nagashima,\altaffilmark{1}
 Tomonori Totani and
 Naoteru Gouda}
\affil{National Astronomical Observatory, Mitaka, Tokyo 181-8588, Japan;
masa@th.nao.ac.jp}
\and
\author{Yuzuru Yoshii}
\affil{Institute of Astronomy, School of Science, The University of
Tokyo, 2-21-1 Osawa, Mitaka, Tokyo 181-8588, Japan}
\affil{Research Center for the Early Universe, Faculty of Science, The
University of Tokyo, Tokyo 113-0033, Japan
}

\begin{abstract}   
 Number counts of galaxies are re-analyzed using a semi-analytic model
 (SAM) of galaxy formation based on the hierarchical clustering
 scenario.  We have determined the astrophysical parameters in the SAM
 that reproduce observations of nearby galaxies, and used them to
 predict the number counts and redshifts of faint galaxies for three
 cosmological models for (1) the standard cold dark matter (CDM)
 universe, (2) a low-density flat universe with nonzero cosmological
 constant, and (3) a low-density open universe with zero cosmological
 constant.  The novelty of our SAM analysis is the inclusion of
 selection effects arising from the cosmological dimming of surface
 brightness of high-redshift galaxies, and also from the absorption of
 visible light by internal dust and intergalactic \ion{H}{1} clouds.
 Contrary to previous SAM analyses which do not take into account such
 selection effects, we find, from comparison with observed counts and
 redshifts of faint galaxies in the Hubble Deep Field (HDF), that the
 standard CDM universe is {\it not} preferred, and a low-density
 universe either with or without cosmological constant is favorable, as
 suggested by other recent studies.  Moreover, we find that a simple
 prescription for the time scale of star formation (SF), being
 proportional to the dynamical time scale of the formation of the
 galactic disk, is unable to reproduce the observed number- redshift
 relation for HDF galaxies, and that the SF time scale should be nearly
 independent of redshift, as suggested by other SAM analyses for the
 formation of quasars and the evolution of damped Ly-$\alpha$ systems.
\end{abstract}   
   
\keywords{cosmology: theory -- galaxies: evolution -- galaxies:
  formation -- large-scale structure of universe }

\section{INTRODUCTION}   

In the field of observational cosmology, counts of the number of faint
galaxies in a given area of sky provide one of the most fundamental
observables from which the cosmological parameters, and hence the
geometry of the Universe, can be determined (e.g., Peebles 1993).  The
predicted number counts of galaxies as a function of apparent magnitude
and redshift, which should be compared with the data, is obtained from
summing up the derived luminosity functions over all redshifts and
morphological types, then multiplied by a redshift-dependent
cosmological volume element.  While the local luminosity function of
galaxies of individual types is known from redshift surveys, the
luminosity function at any high redshift still has to be deduced with
the help of morphological-type-dependent evolution models of galaxies.
Accordingly, the predicted number counts rely directly on how well the
evolution of galaxies is modeled from their formation until the present
(e.g., Yoshii \& Takahara 1988).

With the usual assumption of monolithic collapse, the wind model for
elliptical galaxies and the infall model for spiral galaxies are able to
reproduce many of the observed properties of nearby galaxies, and
provide a strong theoretical tool for understanding their evolution
(Arimoto \& Yoshii 1986, 1987; Arimoto, Yoshii,\& Takahara 1991).
However, when using these traditional evolution models, it has been
claimed that the standard cold dark matter (CDM) universe, or the
Einstein-de Sitter (EdS) universe, is not reconciled with the observed
high counts of faint galaxies, and that a low-density universe is
preferred (Yoshii \& Takahara 1988; Yoshii \& Peterson 1991, 1995;
Yoshii 1993).  Most recently, Totani \& Yoshii (2000, hereafter TY00)
compared their predictions against the number counts observed to the
faint limits in the Hubble Deep Field (HDF, Williams et al. 1996) by
taking into account various selection effects for the first time.
Allowing for the possibility of number evolution of galaxies in a
phenomenological way, and after a comprehensive check of systematic
model uncertainties, they strengthened previous claims and further
demonstrated that mild or negligible merging of high-redshift galaxies
in a low-density flat universe with nonzero cosmological constant is a
likely solution to simultaneously reproduce the observed high counts and
redshift distribution of faint galaxies.  On the other hand, the EdS
universe is in serious contradiction with the data even if a strong
number evolution is invoked.

However, there is a growing evidence from recent observations of the
large-scale structure of the universe that gravitationally bound
objects, such as galaxy clusters, are formed through continuous mergers
of dark halos.  Some authors have tried to construct a galaxy formation
model based on this scenario of hierarchical clustering in a CDM
universe, referred to as a semi-analytic model (SAM) of galaxy formation
(e.g., Kauffmann, White, \& Guiderdoni 1993; Cole et al. 1994).  In
fact, SAMs successfully reproduced a variety of observed features of
local galaxies such as their luminosity function, color distribution,
and so on.

The number counts of faint galaxies have also been analyzed using SAMs
(Cole et al. 1994; Kauffmann, Guiderdoni \& White 1994; Heyl et
al. 1995; Baugh, Cole \& Frenk 1996).  The Durham group claimed that
their model agrees with the observed number counts in the EdS universe
rather than in a low-density universe.  The Munich group reached a
similar conclusion by halting the process of star formation in smaller
halos with circular velocities less than 100 km~s$^{-1}$.  These
conclusions by both groups are apparently at variance with those
obtained using the traditional evolution models of galaxies.  This
apparent discrepancy needs to be immediately resolved.

In this paper we examine the importance of selection effects in
observations of faint galaxies which have been ignored in previous SAM
analyses of galaxy number counts, such as the cosmological dimming of
surface brightness and the absorption of emitted light by internal dust
and intergalactic \ion{H}{1} clouds.  Taking into account these effects
in the same way as in TY00, our SAM analysis obtains, for the first
time, the predicted number counts that can be consistently compared with
the HDF counts to faint magnitude limits.

This paper is outlined as follows.  In \S 2 we briefly describe our SAM.
In \S 3 we constrain the astrophysical parameters in our SAM analysis
against local observations.  In \S 4 we compare theoretical number
counts and redshift distributions of faint galaxies with the data, and
discuss the range of uncertainties in our calculations of galaxy counts.
In \S 5 we provide a summary and discussion.

\section{MODEL}\label{sec:model}

The SAM we employ involves known physical processes connected with the
process of galaxy formation.  It is therefore straightforward to
understand how galaxies form and evolve within the context of this
model.  In the CDM universe, dark matter halos cluster gravitationally
and are merged in a manner that depends on the adopted power spectrum of
initial density fluctuations.  In each of the merged dark halos,
radiative gas cooling, star formation, and gas reheating by supernovae
occur.  The cooled dense gas and stars constitute {\it galaxies}.  These
galaxies sometimes merge together in a common dark halo and more massive
galaxies form.

For the purpose of comparison with observation, we use a stellar
population synthesis approach, from which the luminosities and colors of
model galaxies are calculated.  The SAM well explains the local
luminosity function of galaxies, the color distribution, and so on.  Our
present SAM analysis obtains essentially the same results of previous
SAM analyses, with minor differences in a number of details.  In this
section we only briefly describe our model; its full description will be
given in Nagashima \& Gouda (2001, in preparation).

\subsection{Scheme of Galaxy Formation}\label{sec:scheme}
First, based on the method of Somerville \& Kolatt (1999), we determine
the merging history of dark matter halos by extending the
Press-Schechter formalism (Press \& Schechter 1974; Bower 1991; Bond et
al. 1991; Lacey \& Cole 1993).  We adopt the power spectrum for a
specific cosmology from Bardeen et al.  (1986), and assume a halo with
circular velocity $V_{\rm circ}<$40km~s$^{-1}$ as a diffuse accretion
entity.  The evolution of the baryonic component is followed until the
output redshift with the redshift interval of $\Delta z=0.06(1+z)$,
corresponding to the dynamical time scale of halos which collapse at
that time.  In order to minimize the artificial effect of dividing the
history of galaxy formation into discrete redshift intervals, we fix
$\Delta z=0.06$ until just prior to the output redshift.  This
manipulation is important especially at high redshift.  In our analysis,
the highest output redshift is $z\simeq 9$ with the interval of the
output redshift $\sim$0.05-0.4.

If a dark matter halo has no progenitor halos, the mass fraction of the
gas is given by $\Omega_{\rm b}/\Omega_{\rm 0}$ , where $\Omega_{0}$ is
the density parameter for the total baryonic and non-baryonic
components, and $\Omega_{\rm b}=0.015h^{-2}$ is the baryonic density
parameter constrained by primordial nucleosynthesis calculations (e.g.,
Suzuki, Yoshii, \& Beers 2000).  In this work $h$ represents the Hubble
parameter given by $h=H_{0}/100$ km~s$^{-1}$Mpc$^{-1}$.

When a dark matter halo collapses, the gas in the halo is shock-heated
to the virial temperature of the halo.  We refer to this heated gas as
the {\it hot gas}.  At the same time, the gas in dense regions of the
halo is cooled due to efficient radiative cooling.  We call this cooled
gas the {\it cold gas}.  Assuming an isothermal density distribution of
the entire halo and using the metallicity-dependent cooling function by
Sutherland \& Dopita (1991), we calculate the amount of cold gas which
eventually falls onto a central galaxy in the halo.  In order to avoid
the formation of unphysically large galaxies, the above cooling process
is applied only to halos with $V_{\rm circ}<$500 km~s$^{-1}$ in the
standard CDM and 400 km~s$^{-1}$ in low-density universes.  The physical
reason for this restriction is not clear, but Cole et al. (2000a)
recently reported that the formation of enormous galaxies is hindered in
the isothermal halo having a large core.  Nevertheless, in this paper we
adopt a simple isothermal distribution and prevent the formation of
so-called ``monster galaxies'' by hand.

Stars are formed from the cold gas at a rate of $\dot{M}_{*}={M_{\rm
cold}}/{\tau_{*}}$, where $M_{\rm cold}$ is the mass of cold gas and
$\tau_{*}$ is the time scale of star formation.  We assume that
$\tau_{*}$ is independent of $z$, but dependent on $V_{\rm circ}$ as
follows:

\begin{equation}
\tau_{*}=\tau_{*}^{0}\left(\frac{V_{\rm circ}}{300\mbox{km~s}^{-1}}\right)
^{\alpha_{*}}\label{eqn:taustar}.
\end{equation}

This form is referred to as the ``Durham model'' by Somerville \&
Primack (1999).  It should be noted that Cole et al. (2000a) modified
their original form (equation \ref{eqn:taustar}) as
\begin{equation}
 \tau_*=\tau_*^0\left(\frac{V_{\rm circ}}{200\mbox{km~s}^{-1}}\right)^{\alpha_*}
 \left(\frac{\tau_{\rm dyn}(z)}{\tau_{\rm dyn}(0)}\right),
\label{eqn:taustdyn}
\end{equation}
by multiplying a factor proportional to the dynamical time scale in the
galactic disk.  This form is similar to ``Munich model'' (Kauffman,
White \& Guiderdoni 1993).

For convenience, the cases in equations (\ref{eqn:taustar}) and
(\ref{eqn:taustdyn}) are hereafter referred to as ``constant star
formation (CSF)'' and ``dynamical star formation (DSF),'' respectively.
The free parameters of $\tau_{*}^{0}$ and $\alpha_{*}$ in CSF and DSF
are fixed by matching the observed mass fraction of cold gas in neutral
form in the disks of spiral galaxies. Cole et al. (2000a) stressed that
the $V_{\rm circ}$-dependence is needed to reproduce the observed ratio
of gas mass relative to the $B$-band luminosity of a galaxy.  The effect
of introducing the $\tau_{\rm dyn}$-dependence in DSF will be discussed
in \S \ref{sec:sft}.

In our SAM, stars with masses larger than $10M_\odot$ explode as Type II
supernovae (SNe) and heat up the surrounding cold gas.  This SN feedback
reheats the cold gas to hot gas at rate of $\dot{M}_{\rm reheat}={M_{\rm
cold}}/{\tau_{\rm reheat}}$, where the time scale of reheating is given
by

\begin{equation}
\tau_{\rm reheat}=\left(\frac{V_{\rm circ}}{V_{\rm hot}}
\right)^{\alpha_{\rm hot}} \tau_{*}.
\end{equation}

The free parameters of $V_{\rm hot}$ and $\alpha_{\rm hot}$ 
are determined by matching the local luminosity function of galaxies.  

With these $\dot{M}_{*}$ and $\dot{M}_{\rm reheat}$ thus determined, we
obtain the masses of hot gas, cold gas, and disk stars as a function of
time during the evolution of galaxies.  Chemical enrichment is also
taken into account adopting {\it heavy-element yield} of
$y=0.038=2Z_{\odot}$, but changing this value of $y$ has a minimal
effect on the results described below.

When two or more progenitor halos have merged, the newly formed larger
halo should contain at least two or more galaxies which had originally
resided in the individual progenitor halos.  By definition, we identify
the central galaxy in the new common halo with the central galaxy
contained in the most massive of the progenitor halos.  Other galaxies
are regarded as satellite galaxies.

These satellites merge by either dynamical friction or random collision.
The time scale of merging by dynamical friction is given by
\begin{equation}
\tau_{\rm fric}=\frac{260}{\ln\Lambda_{\rm c}}\left(\frac{R}{\rm Mpc}\right)^{2}
\left(\frac{V_{\rm circ}}{10^{3}{\rm km~s}^{-1}}\right)
\left(\frac{M_{\rm sat}}{10^{12}M_{\odot}}\right)^{-1}{\rm Gyr},
\end{equation}
where $R$ and $V_{\rm circ}$ are the radius and the circular velocity of
the new common halo, respectively, $\ln\Lambda_{\rm c}$ is the Coulomb
logarithm, and $M_{\rm sat}$ is the mass of the satellite galaxies
including the mass of dark matter (Binney \& Tremain 1987).  When the
time elapsed after a galaxy becomes a satellite exceeds $\tau_{\rm
fric}$, a satellite galaxy infalls onto the central galaxy.  On the
other hand, the mean free time scale of random collision is given by
\begin{eqnarray}
\tau_{\rm coll}&=&\frac{500}{N^{2}}\left(\frac{R}{\mbox{Mpc}}
\right)^{3}\left(\frac{r_{\rm gal}}{0.12\mbox{Mpc}}\right)^{-2}
\nonumber\\
&&\qquad\times\left(\frac{\sigma_{\rm gal}}{100\mbox{km~s}^{-1}}
\right)^{-4}\left(\frac{\sigma_{\rm halo}}{300\mbox{km~s}^{-1}}
\right)^{3}\mbox{Gyr},
\end{eqnarray}
where $N$ is the number of satellite galaxies, $r_{\rm gal}$ is their
radius, and $\sigma_{\rm halo}$ and $\sigma_{\rm gal}$ are the 1D
velocity dispersions of the common halo and satellite galaxies,
respectively (Makino \& Hut 1997).  For simplicity, the satellite radius
$r_{\rm gal}$ is set to be one tenth of the virial radius of a
progenitor halo to which the satellite once belonged as a central
galaxy.  With a probability $\Delta t/\tau_{\rm coll}$, where $\Delta t$
is the time step corresponding to the redshift interval $\Delta z$, a
satellite galaxy merges another satellite picked out randomly.  This
process was first introduced in a SAM by Somerville \& Primack (1999).

Consider the case that two galaxies of masses $m_1$ and $m_2 (>m_1)$
merge together.  If the mass ratio $f=m_1/m_2$ is larger than a certain
critical value of $f_{\rm bulge}$, we assume that a starburst occurs and
all the cold gas turns into hot gas, which fills in the halo, and the
stars populate the bulge of a new galaxy.  On the other hand, if
$f<f_{\rm bulge}$, no starburst occurs and a smaller galaxy is simply
absorbed into the disk of a larger galaxy.  These processes are repeated
until the output redshift.

Given the SF rate as a function of time or redshift, the absolute
luminosity and colors of individual galaxies are calculated using a
population synthesis code by Kodama \& Arimoto (1997). The stellar
metallicity grids in the code cover a range of $Z_{*}=$0.0001-0.05. The
initial stellar mass function (IMF) that we adopt is the power-law IMF
of Salpeter form with lower and upper mass limits of $0.1$M$_{\odot}$
and $60$M$_{\odot}$, respectively.  Since our knowledge of the lower
mass limit is incomplete, there is the possibility that many brown
dwarf-like objects are formed.  Therefore, following Cole et al. (1994),
we introduce a parameter defined as $\Upsilon=(M_{\rm lum}+M_{\rm
BD})/M_{\rm lum}$, where $M_{\rm lum}$ is the total mass of luminous
stars with $m\geq 0.1M_\odot$ and $M_{\rm BD}$ is that of invisible
brown dwarfs.

To account for extinction by internal dust we adopt a simple model by
Wang \& Heckman (1996) in which the optical depth in $B$-band is related
to the luminosity as $\tau_{B}=0.8(L_{B}/1.3\times
10^{10}L_{\sun})^{0.5}$.  Optical depths in other bands are calculated
by using the Galactic extinction curve, and the dust distribution in
disks is assumed to be the slab model considered by Somerville \&
Primack (1999).  It is not trivial that such an empirical dust model can
be extraporated to very bright galaxies.  However, since the extinction
is typically $A_{B}\sim 1$ mag at the bright-end of the local luminosity
function, our results especially in the $I$-band are not significantly
affected by the details of the dust model.

Emitted light from distant galaxies is absorbed by Lyman lines and Lyman
continuum in intervening intergalactic \ion{H}{1} clouds.  The redshift
at which this effect becomes important is different for different
photometric passbands, because the absorption occurs around 1000 \AA\ in
the rest frame of the clouds.  Fig. \ref{fig:h1clouds} shows the
expected extinction as a function of redshift for four passband filters
of the HDF (Yoshii \& Peterson 1994).  This effect has been included
into a SAM by Baugh et al. (1998) in order to pick out the Lyman-break
galaxies at high redshift by color selection criteria.

We classify galaxies into different morphological types according to the
$B$-band bulge-to-disk luminosity ratio $B/D$.  In this paper, following
Simien \& de Vaucouleurs (1986), galaxies with $B/D\geq 1.52$, $0.68\leq
B/D<1.52$, and $B/D<0.68$ are classified as ellipticals, S0s, and
spirals, respectively.  Kauffmann et al.  (1993) and Baugh, Cole \&
Frenk (1996) showed that this method of type classification well
reproduces the observed type mix.

The above procedure is a standard one in the SAM of galaxy formation.
In the next subsection, in order to investigate the properties of
high-$z$ galaxies properly, we introduce two important effects into our
SAM analysis.

\subsection{New ingredients of the model: selection effects}
\label{sec:luminosity}

We judge whether the surface brightness of the galaxies in our SAM is
above the detection threshold of the HDF observations.  The intrinsic
size of spiral or disk-dominated galaxies is estimated by adopting the
dimensionless spin parameter $\lambda_{\rm H}=0.05$, and by conserving
the specific angular momentum during the gas cooling.  On the other
hand, the intrinsic size of early-type galaxies is estimated from their
virial radii, adjusted by a scaling parameter $f_{\rm b}$ to match with
the observed size of early-type galaxies.  Note that the selection
effects on the local luminosity function have been considered by Cole et
al. (2000a) in a simple way in which the isophotal magnitude within 25
mag arcsec$^{-2}$ was used.

The selection effects in predicting the HST number counts in our SAM
analysis are evaluated as follows.  Using the intrinsic size of model
galaxies as obtained above, and adopting a Gaussian point-spread
function $f(x)$ for the HST observations, the surface brightness profile
$\tilde{g}(x)$ in the observer frame is given by
\begin{equation}
\tilde{g}(\vert\bx\vert)=\int{\rm d}\bx' 
f(\vert\bx'-\bx\vert)g(\vert\bx'\vert),
\end{equation}
where $\vert\bx\vert=x=r/r_{\rm e}$ is the normalized radius away from 
the galaxy center relative to the effective radius $r_{\rm e}$, and 
$g(x)=\exp(-a_nx^{1/n})$ is the intrinsic surface brightness profile.  
We adopt $n=1$ for spirals and $n=4$ for ellipticals. The coefficient 
$a_n$ is given by $a_1=1.68$ and $a_4=7.67$. Then, for each of our model
galaxies, we determine their surface brightness profile $S(\theta)$, 
where $\theta=xr_{\rm e}/d_{\rm A}$ and $d_{\rm A}$ is the 
angular-diameter distance.  We note that model galaxies with surface 
brightness brighter than the threshold $S_{\rm th}$ and with an 
isophotal diameter larger than the minimum diameter $D_{min}$ are 
actually detected as galaxies (Yoshii 1993). In order to be consistent 
with the HDF observations, we use the isophotal magnitude scheme, 
$S_{\rm th}=27.5$ mag~arcsec$^{-2}$ in $V_{606}$, $S_{\rm th}=27.0$ 
mag~arcsec$^{-2}$ in $I_{814}, U_{300}$ and $B_{450}$, and $D_{min}\sim$0.2 
arcsec.  More details are described in TY00.

\section{THE SETTING OF PARAMETERS IN OUR SAM ANALYSIS}\label{sec:norm}

We consider the predicted number counts in four models -- SC, OC, LC,
and LD (Table \ref{tab:astro}).  The first three models are for CSF
(equation \ref{eqn:taustar}), and the fourth model is for DSF (equation
\ref{eqn:taustdyn}).  The capitals S, O, and L refer to the standard CDM
universe, a low-density open universe and a low-density flat universe
with non-zero cosmological constant ($\Lambda$), respectively.

The cosmological parameters ($\Omega_0$, $\Omega_\Lambda$, $h$,
$\sigma_8$) are tabulated in Table \ref{tab:astro}.  For all the models
the baryon density parameter $\Omega_{\rm b}=0.015h^{-2}$ is used in
common.  For the low-density open and flat universes the value of
$\sigma_8$ is determined from observed cluster abundances (Eke, Cole \&
Frenk 1996).

The astrophysical parameters ($V_{\rm hot}$, $\alpha_{\rm hot}$, $\tau_*^0$,
$\alpha_*$, $f_{\rm b}$, $f_{\rm bulge}$, $\Upsilon$) are constrained from
local observations.  However, since the parameter of $f_{\rm bulge}$, among
others, do not affect our result, we set $f_{\rm bulge}=0.2$ for the standard
CDM universe, and $f_{\rm bulge}=0.5$ for the low-density open and flat
universes.  Other parameters are discussed below.

\subsection{The Local Luminosity Function of Galaxies}{\label{sec:lf}}  

The SN feedback-related parameters of $V_{\rm hot}$ and $\alpha_{\rm
hot}$ determine the location of the knee of the luminosity function and
the faint-end slope, respectively.  It should be noted that the mass
fraction $\Upsilon$ of invisible stars determines the magnitude scale of
galaxies, so that changing $\Upsilon$ moves the luminosity function
horizontally without changing its overall shape.  Therefore, coupled
with $V_{\rm hot}$, $\Upsilon$ determines the bright portion of the
luminosity function.

Fig. \ref{fig:lf} shows theoretical results represented by thick lines
for the four models tabulated in Table \ref{tab:astro}.  Symbols with
errorbars indicate observational results from the $B$-band redshift
surveys such as APM (Loveday et al. 1992), ESP (Zucca et al. 1997),
Durham/UKST (Ratcliffe et al.  1997), 2dF (Folkes et al. 1999) and SDSS
(Blanton et al. 2000), and from the $K$-band redshift surveys (Mobasher
et al. 1993; Gardner et al.  1997; 2MASS, Cole et al. 2000b).  It is
evident that while most of $B$-band redshift surveys give a rather flat
slope in the faint end, the ESP and SDSS surveys give an extreme case
showing a much steeper slope.  Note that the SDSS luminosity function
shown in Fig. \ref{fig:lf} is that with the same detection limit as
employed in the 2dF survey.

In this paper we have chosen the values of SN feedback-related
parameters so as to reproduce a flat faint-end slope, but in
\S\ref{sec:fb} we will investigate the effect of using the steepest ESP
slope in predicting the number counts of faint galaxies.

\subsection{The Mass Fraction of Cold Gas in Spiral Galaxies}{\label{sec:gas}} 

The SF rate-related parameters of $\tau_*^0$ and $\alpha_*$ determine
the overall mass fraction of cold gas in galaxies with given luminosity
and its luminosity dependence, respectively.  In some previous SAM
analyses, the mass fraction of gas in the Milky Way is exclusively used
to fix $\tau_*^0$ for all other spiral galaxies. Then, when
$\alpha_*=0$, Cole et al. (2000a) found that dwarf-size galaxies have too
little gas to be consistent with observations.  According to Cole et
al. (2000a), we here constrain both $\tau_*^0$ and $\alpha_*$ in a
combined manner to reproduce the mass fraction of cold gas in galaxies
spanning a wide range of luminosity.

Fig. \ref{fig:gas} shows the ratio of cold gas mass relative to
$B$-band luminosity of spiral galaxies as a function of their
luminosity.  Solid curves show theoretical results for the four models
tabulated in Table \ref{tab:astro}.  We here assume that 75\% of the
cold gas in these models is comprised of hydrogen, i.e., $M_{\rm
HI}=0.75M_{\rm cold}$.  Filled diamonds with errorbars indicate the
\ion{H}{1} data taken from Huchtmeier \& Richter (1988).

It should be noted that use of $\tau_*^0$ for either CSF (equation
\ref{eqn:taustar}) or DSF (equation \ref{eqn:taustdyn}) hardly affects
the resulting luminosity function, as seen from the difference between
LC and LD in Fig. \ref{fig:lf}.  However, $\alpha_*$ depends on the
strength of the SN feedback, and the strength should be different for
different universe models (see Table \ref{tab:astro}).  Therefore, in
the case of $\alpha_*$, we can only fix its value after the SN
feedback-related parameters are constrained in a specified universe
model.

\subsection{The Intrinsic Sizes of Galaxies}{\label{sec:size}}

The scaling parameter $f_{\rm b}$ is determined by matching the
effective radius of early-type $L_*$ galaxies, while the actual
$L$-dependence in $f_{\rm b}$ is ignored for simplicity.  We discuss the
effect of this assumption in \S \ref{sec:sizedis}.

Knowledge of the intrinsic size of galaxies is a necessary quantity in
our SAM analysis, because it is the surface brightness of galaxies, not
their luminosity, that is relevant to evaluating the selection effects
for model galaxies.  For spiral galaxies, the key process is the
conservation of specific angular momentum during the cooling of hot gas.
The effective disk radius $r_{\rm e}$ is given by $r_{\rm
e}=(1.68/\sqrt{2})\lambda_{\rm H}r_{\rm i}$, where $r_{\rm i}$ is an
initial radius of the progenitor gas sphere (Fall 1983).  However, the
definition of disk size during the cooling phase is not readily apparent
from this formula.  Furthermore, the angular momentum transfer during
mergers of galaxies in a merged halo is very complicated.  Without
entering into all these complexities, in this paper we re-estimate the
disk size by the above equation when the disk mass increases twice.
This approach is simple but reproduces the observations rather well.

Fig. \ref{fig:rad} shows the effective disk radii of spiral galaxies
as a function of their luminosity.  Thick solid lines show the
theoretical results for the four models tabulated in Table
\ref{tab:astro}.  Dotted lines indicate the best-fit relation to the
observational data given by Impey et al. (1996).  All models provide
reasonable disk sizes for dwarf galaxies.  Slight deviations from the
observations of bright galaxies may have occurred partly because dust
obscuration becomes effective for such galaxies (cf. \S
\ref{sec:luminosity}), and partly because our method of defining the
disk size is too simple.  We discuss the effect of changing the disk
size in \S \ref{sec:sizedis}.

In this paper, the effective radius of elliptical galaxies is estimated
from scaling the virial radius by a parameter $f_{\rm b}$, i.e., $r_{\rm
e}=f_{\rm b}GM_{\rm b}/V_{\rm circ}^2$, where $M_{\rm b}$ is the mass of
stars and cold gas, and $f_{\rm b}$ is fixed to reproduce the effective
radius of an $L_*$ galaxy.  In Fig. \ref{fig:rad}, thick dashed lines
indicate the effective radius of elliptical galaxies for the four models
tabulated in Table \ref{tab:astro}.  Dashed and dot-dashed lines show
the best-fit relations for dwarf and compact ellipticals, respectively,
based on the observational data given by Bender et al. (1992).
Unfortunately, our SAM analysis cannot reproduce two distinct sequences
simultaneously; the theoretical $r_{\rm e}-L_B$ relation becomes
coincident with the bright portion of the giant-dwarf sequence, whereas
the same relation changes its slope becoming coincident with the faint
portion of giant-compact sequence.  We find that this changeover
magnitude is determined by the strength of SN feedback (see Appendix
\ref{sec:app}), and that the extreme manipulation that reproduce the
dwarf sequence affects the number counts of model galaxies only weakly.

\section{Results}
\subsection{Galaxy Number Counts}\label{sec:counts}

\begin{figure*}
\epsfxsize=14cm
\epsfbox{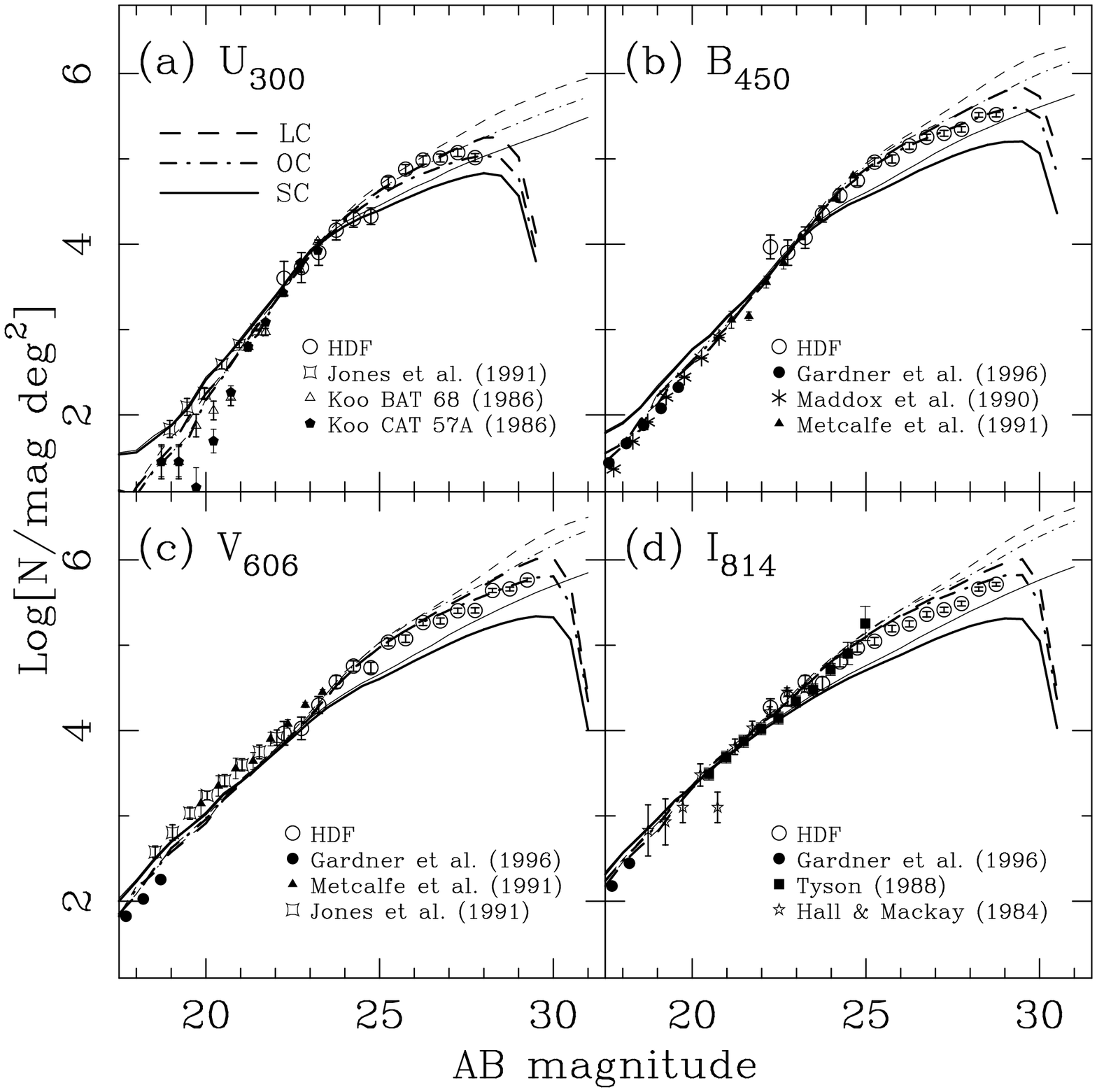}
\caption{Number-magnitude relations for various cosmological models.
 The solid, dot-dashed, and dashed lines indicate SC, OC, and LC,
 respectively.  The thick lines denote models including the selection
 effects and the absorption effect by intergalactic \ion{H}{1} clouds,
 while the thin lines denote those without the effects.  The symbols
 indicate observational data.  }
\label{fig:counts}
\end{figure*}

Fig. \ref{fig:counts} shows the number counts of galaxies as a
function of apparent isophotal magnitude for the SC, OC, and LC models
in the HST ${UBVI}$ bands in the AB magnitude system.  The thick lines
are the theoretical predictions, based on the HST observational
conditions, including the selection effects from the cosmological
dimming of surface brightness and also from the absorption of visible
light by internal dust and intergalactic \ion{H}{1} clouds.  The thin
lines are the predictions ignoring the selection effects except for the
effect of dust absorption, because the dust absorption is already taken
into account in reproducing the local luminosity function.  Open circles
with errorbars show the HDF data (Williams et al.  1996), and other
symbols show ground-based data after transformation to AB magnitudes to
be consistent with the HST.

In order to discriminate favorable universe models, the predictions
shown by the thick lines should be compared directly with the
observational data.  It is evident from this figure that the SC model
falls short of the HST data.  We note that the discrepancy between the
SC model and the data is seen at $B_{450}\gtrsim 25$, even if we do not
consider the selection effects.  On the other hand, both of the LC and
OC models agree well with the data, owing to the inclusion of selection
effects.  We will discuss the uncertainty of estimating these effects in
the next section.

Fig. \ref{fig:zdist} shows the redshift distribution of galaxies.  The
thick lines denote the SC, OC and LC models, including all the selection
effects as in Fig. \ref{fig:counts}.  The histogram indicates the
observed redshift distribution based on photometric redshifts of HDF
galaxies estimated by Furusawa et al. (2000), in which they improved the
method of redshift estimation compared to the method by
Fern{\'a}ndez-Soto, Lanzetta \& Yahil (1999).

In the LC model, both the peak height and the distribution towards
higher redshift agree well with the observation.  However, the peak
height of the SC model falls significantly short of that observed at
$z\simeq 1-1.5$. This is a direct reflection of the lack of galaxies in
the theoretical number-magnitude relation for the SC model (see
Fig.\ref{fig:counts}).  In the OC model, the relative number between the
peak and high-$z$ tail at $z\gtrsim 3$ is smaller than that observed and
is inconsistent with the data.  This is a direct reflection of the
$z$-dependence of the comoving volume element $dV/dz$, which is shown in
the middle panel of Fig.\ref{fig:zdist}.  While in the LC model the
matter density dominates over the cosmological constant at high
redshift, which leads to a similar $z$-dependence of $dV/dz$ to that in
the EdS universe, the negative curvature effect makes $dV/dz$ decline
more slowly at such high redshift in the OC model.  Thus the redshift
distribution in the OC model declines toward higher redshift more slowly
than that in the LC model.  In order for the OC model to agree with the
data, we would need to halt the star formation at $z\gtrsim 3$ by hand,
for unknown reasons.

We conclude that as far as that the astrophysical parameters in our SAM
analysis are constrained by local luminosity function of galaxies, the
standard CDM universe does not agree with both the observed
number-magnitude and number-redshift relations, and that the
$\Lambda$-dominated flat universe is best able to reproduce these
relations simultaneously.

\subsection{Uncertainties in the SAM counts}{\label{sec:discussion}}

In this section we discuss uncertainties in predicting the number counts
of galaxies.  Sources of such uncertainties considered here include the
time scale of star formation (\S \ref{sec:sft}), the galaxy size (\S
\ref{sec:sizedis}), and the adopted SN feedback (\S \ref{sec:fb}).

\subsubsection{Star Formation Time-Scale}\label{sec:sft}

\begin{figure*}
\epsfxsize=14cm
\epsfbox{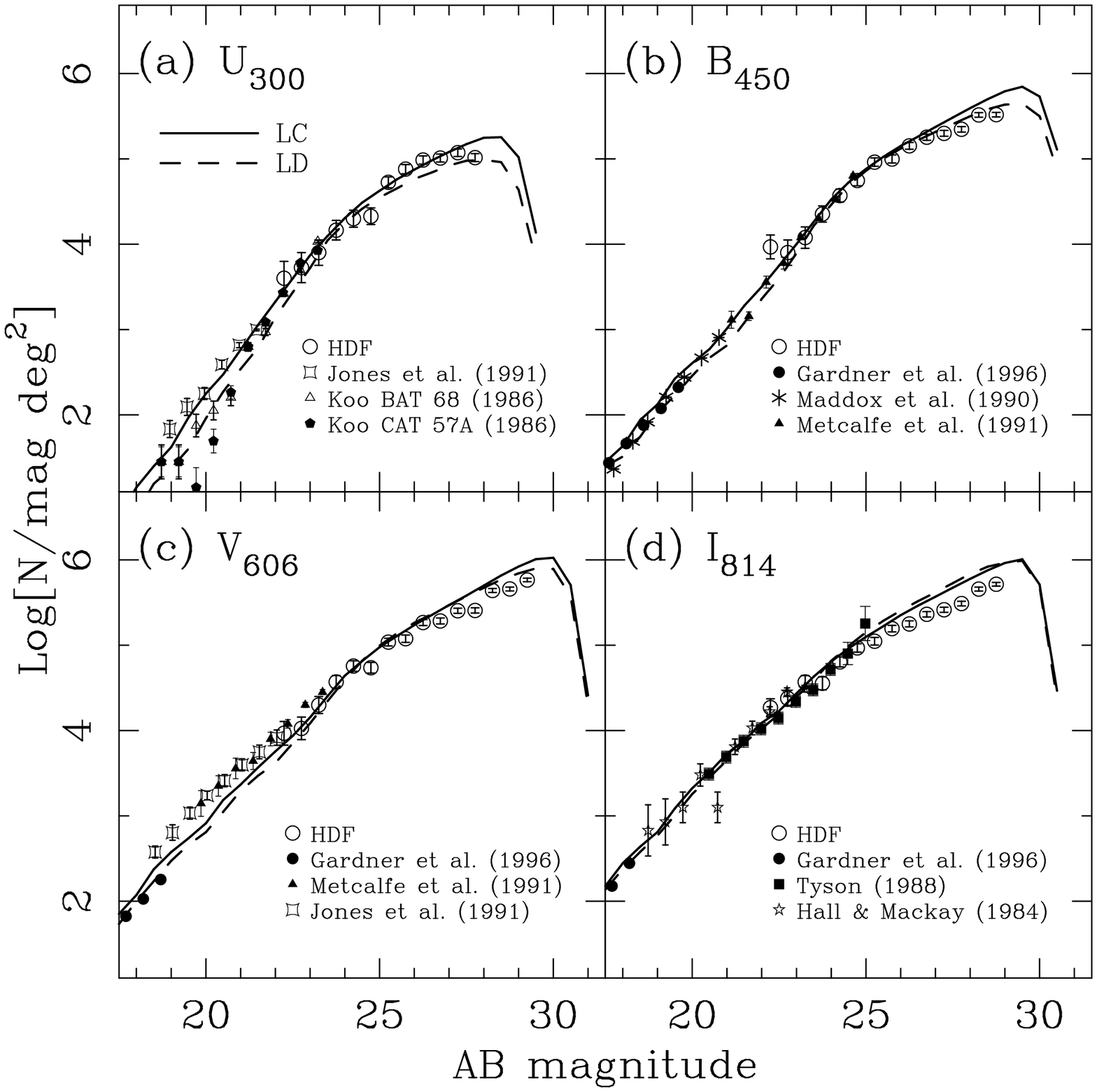}
\caption{Number-magnitude relations.  The solid lines and the dashed
lines indicate the LC (CSF) and LD (DSF) models, respectively.  }
\label{fig:sft}
\end{figure*}

Here we evaluate the effects of changing the time-scale of star
formation from CSF (equation \ref{eqn:taustar}) to DSF (equation
\ref{eqn:taustdyn}).  Fig. \ref{fig:sft} shows predicted number counts
for the corresponding $\Lambda$-dominated LC and LD models.  The
difference between these models is apparently small but its
magnitude-dependence is different among the predictions in the $UBVI$
bands.  In the case of longer wavelength such as the $I_{814}$ band, the
number of faint galaxies in the LD model becomes larger than that in the
LC model, because in the case of DSF more stars are formed at high
redshift according to much a shorter $\tau_*({\rm DSF})$ as compared
with CSF.  [In the particular case of the EdS universe, not shown here,
the DSF gives $\tau_*({\rm DSF})\propto (1+z)^{-3/2}$, so that the
present $\tau_*({\rm DSF})$ is more than 10 times longer than that at
$z\sim 5$.]  However, in the case of shorter wavelengths, particularly
the $U_{300}$ band, because the apparent luminosity from galaxies is
dominated by instantaneous SF, and because many stars have already been
formed at higher redshift, the number of faint galaxies in the LD model
becomes slightly smaller than that in the LC model.

The difference between the LC and LD models is most prominent when we
consider the redshift distribution of faint galaxies.  Fig.
\ref{fig:z_sft} shows such predictions for the LC and LD models.
Clearly the LD model predicts too many high-$z$ galaxies to be
consistent with observation.  We may be able to remedy this defect by
imposing a more efficient internal dust absorption in order to decrease
the number of high-$z$ galaxies.  However, this manipulation simply
decreases the number of faint galaxies below that observed in the
$U_{300}$.  Thus, we suggest that the $z$-dependence of SF should be
negligible and the CSF is a reasonable option in the framework of our
SAM.

\subsubsection{Galaxy Size}\label{sec:sizedis}

\begin{figure*}
\epsfxsize=14cm
\epsfbox{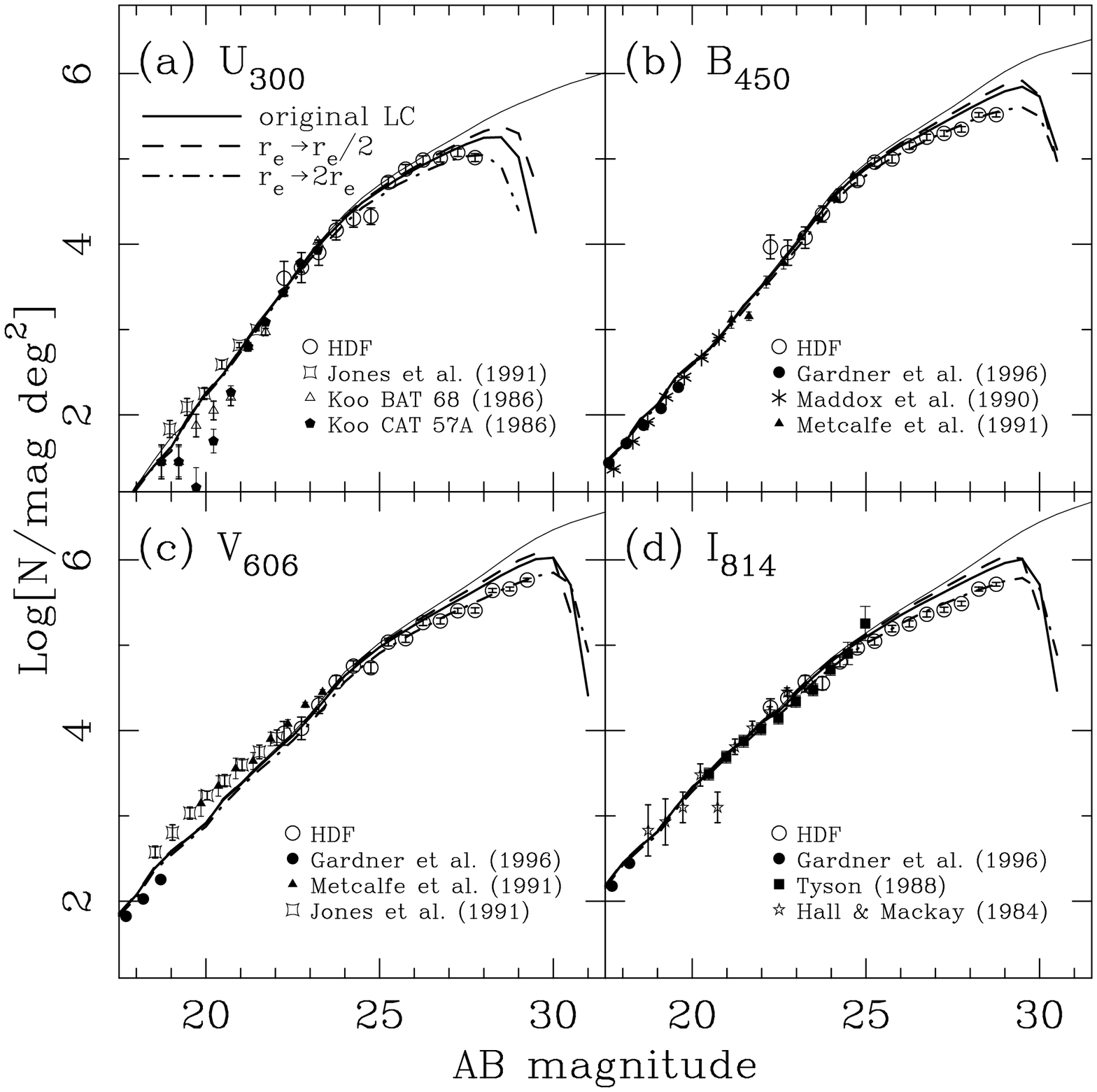}
\caption{ Number-magnitude relations.  The solid lines indicate the
original LC model.  The thick and thin lines denote models with and
without the selection effects.  The dashed lines and the dot-dashed
lines indicate the high and low surface brightness models, respectively.
} \label{fig:rad_l}
\end{figure*}

In this paper, it is assumed according to the usual SAM analysis that
the physical mechanism for determining the galaxy size is the
conservation of specific angular momentum for spiral galaxies and the
virial theorem for the baryon component of elliptical galaxies.  While
this assumption is considered to be reasonable, it is important to note
the uncertainties in the normalization of the above relations.  In this
section we consider two LC variants and compare them with the original
LC model.  Figs. \ref{fig:rad_l} and \ref{fig:z_rad_lcdm} present
theoretical predictions from these three LC models.  We rather
arbitrarily decrease and increase the original galaxy size by a factor
of 2, and refer to these variants as ``high surface brightness'' and
``low surface brightness,'' respectively.  Note that the 1$\sigma$
scatter in the observational data is about a factor of 1.7, so the range
of changing the radius is sufficient large to check the uncertainty in
the galaxy size.  Given the threshold of surface brightness for
detection used in the HST observation, model galaxies with ``high
surface brightness'' are more easily detected, while those with ``low
surface brightness'' remain undetected.  The number counts of these two
variants differ only by a factor of 1.5 at $B_{450}\sim 28$, as well as
at the corresponding magnitudes in the other bands.  We found that the
uncertainties from changing the galaxy size in the SC and OC models are
almost the same as those in the LC model.

\subsubsection{Supernova Feedback}\label{sec:fb}
The SN feedback-related parameters essentially determine the resulting
shape of the local luminosity function of galaxies (\S \ref{sec:lf}).
Therefore, changing these parameters corresponds to changing the number
density of galaxies in the local universe.  In this section, we
constrain these parameters against the ESP luminosity function (Zucca et
al. 1997) which gives the steepest faint-end slope among other
observations.  This is done by weakening the strength of SN feedback in
three CSF variants such as SCw, OCw, and LCw.  Other parameters such as
the SF rate-related are determined by the same way in \S\ref{sec:norm}.
All the parameters for these models are tabulated in Table
\ref{tab:astro2}, and their $B$-band luminosity functions are shown in
Fig.  \ref{fig:lf_fb}.

\begin{figure*}
\epsfxsize=14cm
\epsfbox{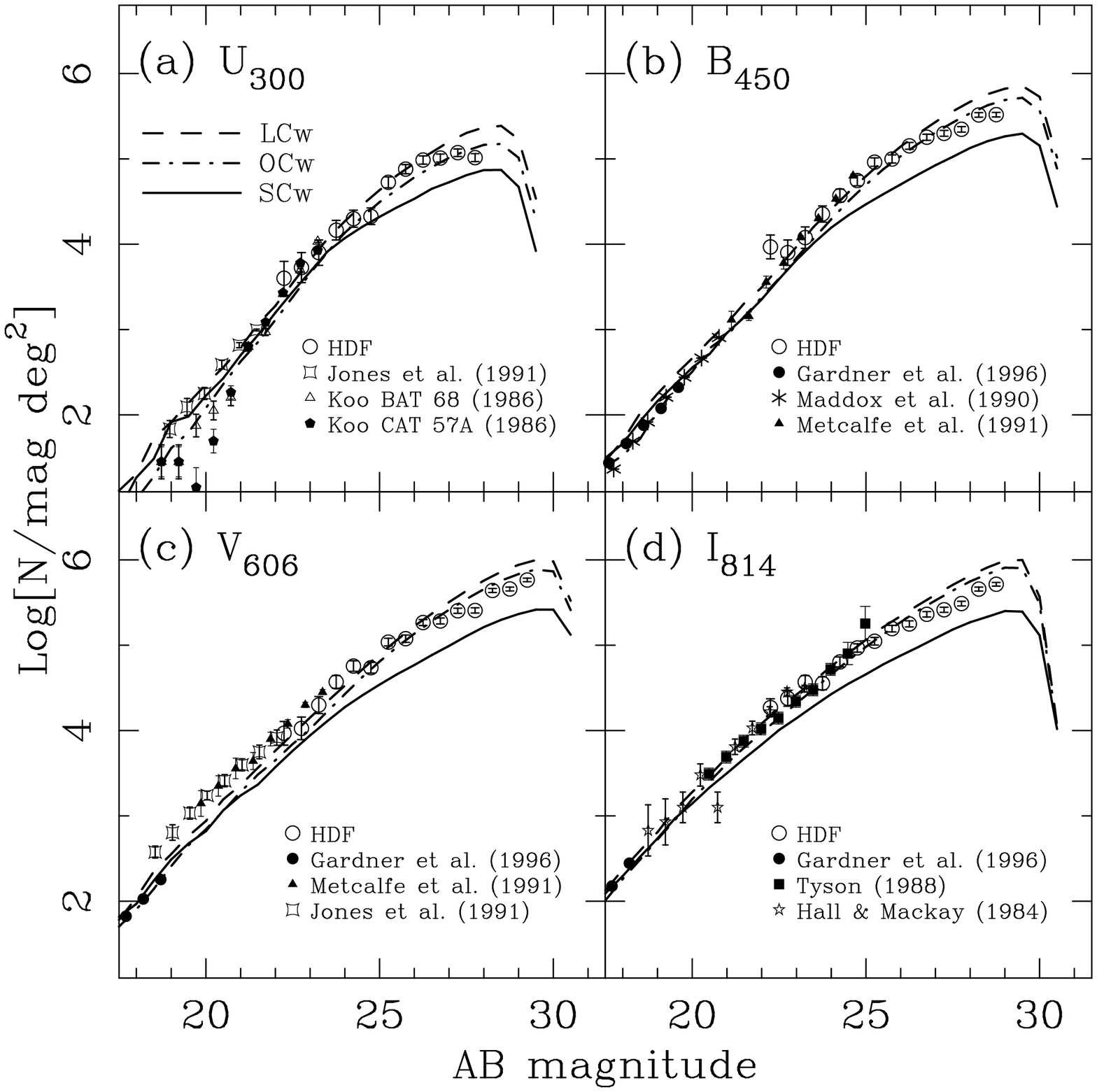}
\caption{Number-magnitude relations.  The same as Fig.
\ref{fig:counts} but for the model SCw, OCw and LCw.  The types of lines
are the same as Fig. \ref{fig:lf_fb}.}
\label{fig:fb}
\end{figure*}

Straightforward calculations predict the number counts of galaxies which
exceed the bright counts at $B\sim 20$, and this excess alone may
invalidate the extreme assumption for faint-end slope of the local
luminosity function.  However, considering the possible existence of
systematic uncertainties that could affect current observations of the
local luminosity function, we adjust the normalization in such a way as
to reproduce the observed bright counts.  Then we examine the effect of
adopting the steepest ESP slope instead of our standard choice.  Fig.
\ref{fig:fb} shows the number counts of galaxies for the SCw, OCw, and
LCw models.  It is evident from this figure that even the steepest ESP
slope does not save the standard CDM universe, which confirms the claim
by TY00.

\section{SUMMARY AND DISCUSSION}{\label{sec:conclusion}}

We have calculated the number counts of faint galaxies in the framework
of a SAM for three cosmological models of the standard CDM (EdS)
universe, a low-density open universe, and a low-density flat universe
with nonzero $\Lambda$.  The novelty of our SAM analysis is that
theoretical predictions are made by fully taking into account the
selection effects from the cosmological dimming of surface brightness of
galaxies and also from the absorption of visible light by internal dust
and intergalactic \ion{H}{1} clouds.

Comparison of theoretical predictions with the observed number counts
and photometric redshift distribution of HST galaxies, as well as other
ground-based observations, indicates that the standard CDM is ruled out
and a $\Lambda$-dominated flat universe is most favorable, while a
low-density open universe is marginally favored. This result is in sharp
contrast with previous SAM analyses on galaxy number counts where many
of the conceivable selection effects in faint observations have been
ignored.  It is only recently that the SAM analyses have included the
effects of internal dust absorption (Somerville \& Primack 1999; Cole et
al. 2000a), and intergalactic \ion{H}{1} absorption (Baugh et al. 1998),
and the isophotal selection effect in a very simple way (Cole et
al. 2000a).  However, as stressed by TY00, any predictions based on
number count analyses will be seriously compromised unless all of the
selection effects considered in this paper are taken into account
simultaneously.

Based on a hierarchical clustering scenario in the CDM universe, the SAM
naturally involves the merger-driven number evolution of galaxies, which
has been introduced only phenomenologically in traditional models of
galaxy evolution (Yoshii 1993; Yoshii \& Peterson 1995; TY00). The fact
that we have essentially reached the same conclusion from
phenomenological approach confirms that previous simple prescriptions of
the number evolution of galaxies are still useful in studying the global
evolution of faint galaxies.  It should be noted that in TY00 the
adopted number evolution law which reproduces the observational data is
$\phi_{*}\propto (1+z)$ and $L_{*}\propto (1+z)^{-1}$, and is consistent
with the observational estimate of the merger rate of the Canada-France
redshift survey (CFRS) galaxies (Le F{\'e}vre et al. 2000).

The basic ingredients in the SAM analysis include the SF process and SN
feedback.  Although the constant and dynamical SFs can equally reproduce
the local luminosity function by adjustment of their free parameters, we
find that the dynamical SF predicts the formation of too many high-$z$
galaxies to be consistent with the photometric redshift distribution of
faint HST galaxies.  Thus, our SAM analysis prefers the constant SF, as
is also supported from other recent SAM analyses on the formation of
quasars (Kauffmann \& Haehnelt 2000) and the evolution of damped
Ly-$\alpha$ systems (Somerville, Primack \& Faber 2001).
  
The SN feedback, associated with the virial equilibrium for the baryonic
component, controls the resulting size versus luminosity relation of
elliptical galaxies.  However, our SAM analysis is unable to reproduce
the observed relation bifurcating into the giant-dwarf and giant-compact
sequences; our brighter ellipticals reside on the bright portion of
giant-dwarf sequence and our fainter ellipticals on the faint portion of
giant-compact sequence.  Although this limitation of our SAM analysis is
found to hardly affect the conclusion in this paper, it is of urgent
importance for the SAM to be equipped with some mechanism enabling the
bifurcation of early-type galaxies into two distinct sequences as
observed.

\appendix
\section{BULGE SIZE AND SUPERNOVA FEEDBACK}\label{sec:app}

In \S\ref{sec:sizedis} we mentioned that the intrinsic size of
early-type galaxies is related to the strength of SN feedback.  In \S2.1
describing SN re-heating the baryonic mass in a halo is given by $M_{\rm
b}\lesssim \Omega_{\rm b}M_{\rm H}/(1+\beta)$, where $\beta\equiv
\tau_*/\tau_{\rm reheat}$ and $M_{\rm H}$ is the halo mass.  The left-
and right-hand sides are not equal to each other, because this relation
depends on the merging history of halos.  The spherical collapse model
gives $M_{\rm H}\propto V_{\rm circ}^3$, ignoring the dependence on the
formation redshift.  Thus, we obtain $r_{\rm e}\propto M_{\rm b}/V_{\rm
circ}^2\propto V_{\rm circ}/(1+\beta)$.  Assuming a constant ratio of
baryonic mass $M_{\rm b}$ relative to luminosity $L$, two limiting cases
of $\beta\ll 1$ and $\beta\gg 1$ give
\begin{equation}
 \log r_{\rm e}=\left\{
\begin{array}{ll}
 \displaystyle{-\frac{1}{7.5}{\cal M}}+\mbox{const.}& \mbox{for~~~} 
\beta\ll 1\\
 \displaystyle{-\frac{(1+\alpha_{\rm hot})}{2.5(3+\alpha_{\rm hot})}{\cal 
M}}+\mbox{const.}& \mbox{for~~~} \beta\gg 1,\\
\end{array}\right.\label{eqn:rad}
\end{equation}
where $\cal M$ is the absolute magnitude.  Fig. \ref{fig:rad_ana}
shows the size of early-type galaxies for the SC model ({\it upper thick
curve}) and the LC model ({\it lower thick curve}).  The vertical scale
is chosen arbitrarily to avoid overlapping.  Solid lines are for
$\beta\ll 1$ in equation (\ref{eqn:rad}).  Dot-dashed and dashed lines
are for $\beta\gg 1$ with $\alpha_{\rm hot}=5.5$ and 2.5, respectively
(see Table \ref{tab:astro}).  Although we fixed the baryon fraction, the
baryonic mass-to-light ratio, and the formation redshift of halos,
theoretical results ({\it thick curves}) can readily fit the scaling
relations.  This indicates that the SN feedback process essentially
determines the size of early-type galaxies.  In other words, we can
constrain the SN feedback from reproducing their observed size versus
luminosity relation.  It should be noted that the importance of SN
feedback in galaxy evolution has also been highlighted in other recent
SAM analyses of the color versus magnitude relation of early-type
galaxies (Kauffmann \& Charlot 1998; Nagashima \& Gouda 1999).

\acknowledgments

We would like to thank T. C. Beers for his critical reading of the
manuscript.  This work has been supported in part by the Grant-in-Aid
for the Center-of-Excellence research (07CE2002) and for the Scientific
Research Funds (10640229 and 12047233) of the Ministry of Education,
Science, Sports and Culture of Japan.

\twocolumn
\begin{inlinefigure}
\includegraphics[width=8cm]{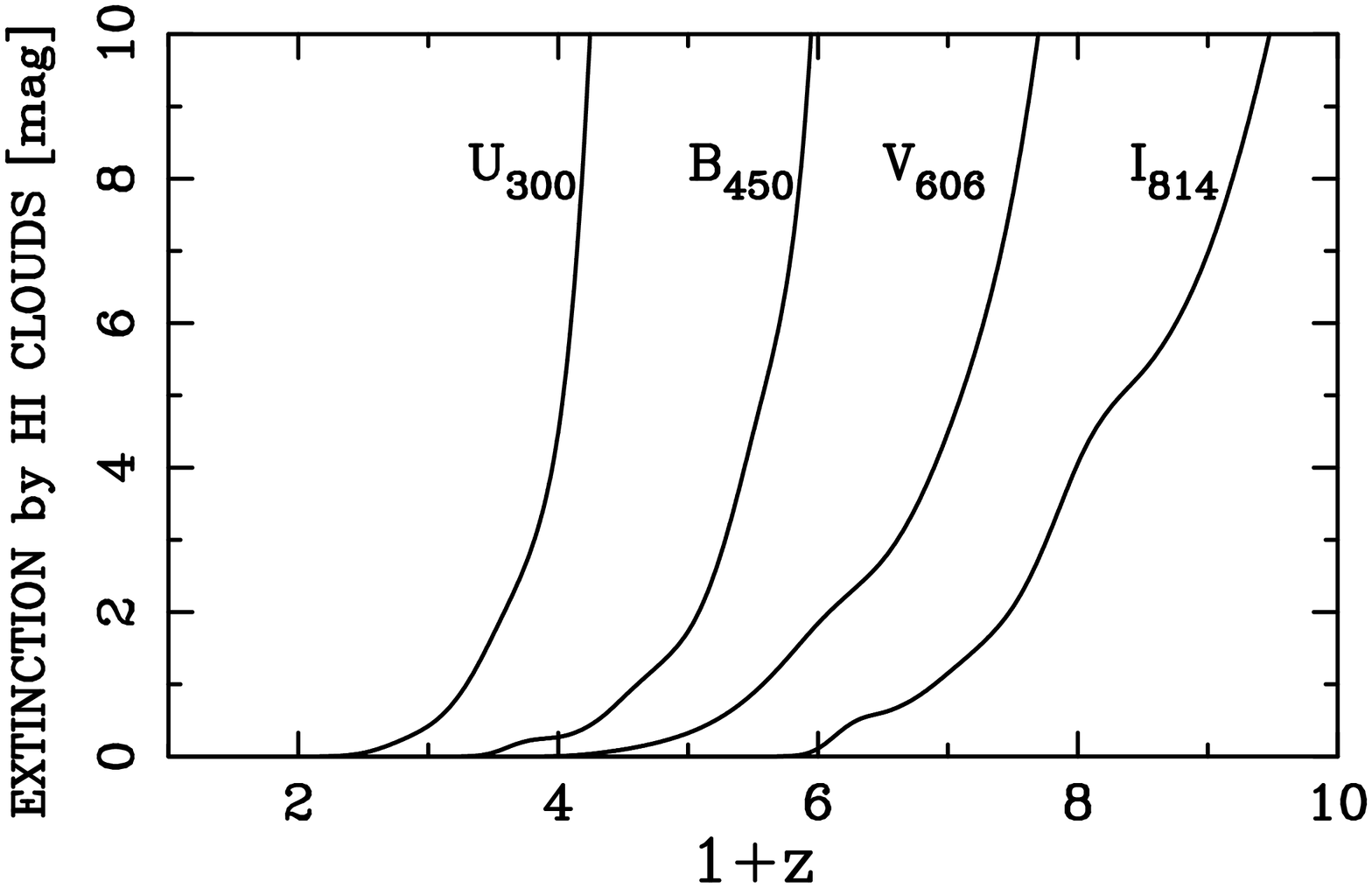}
\caption{Extinction in units of magnitude by intervening intergalactic 
\ion{H}{1} clouds for four HDF passband filters.  The model optical depth 
is taken from Yoshii \& Peterson (1994).  }
\label{fig:h1clouds}
\end{inlinefigure}

\begin{inlinefigure}
\includegraphics[width=8cm]{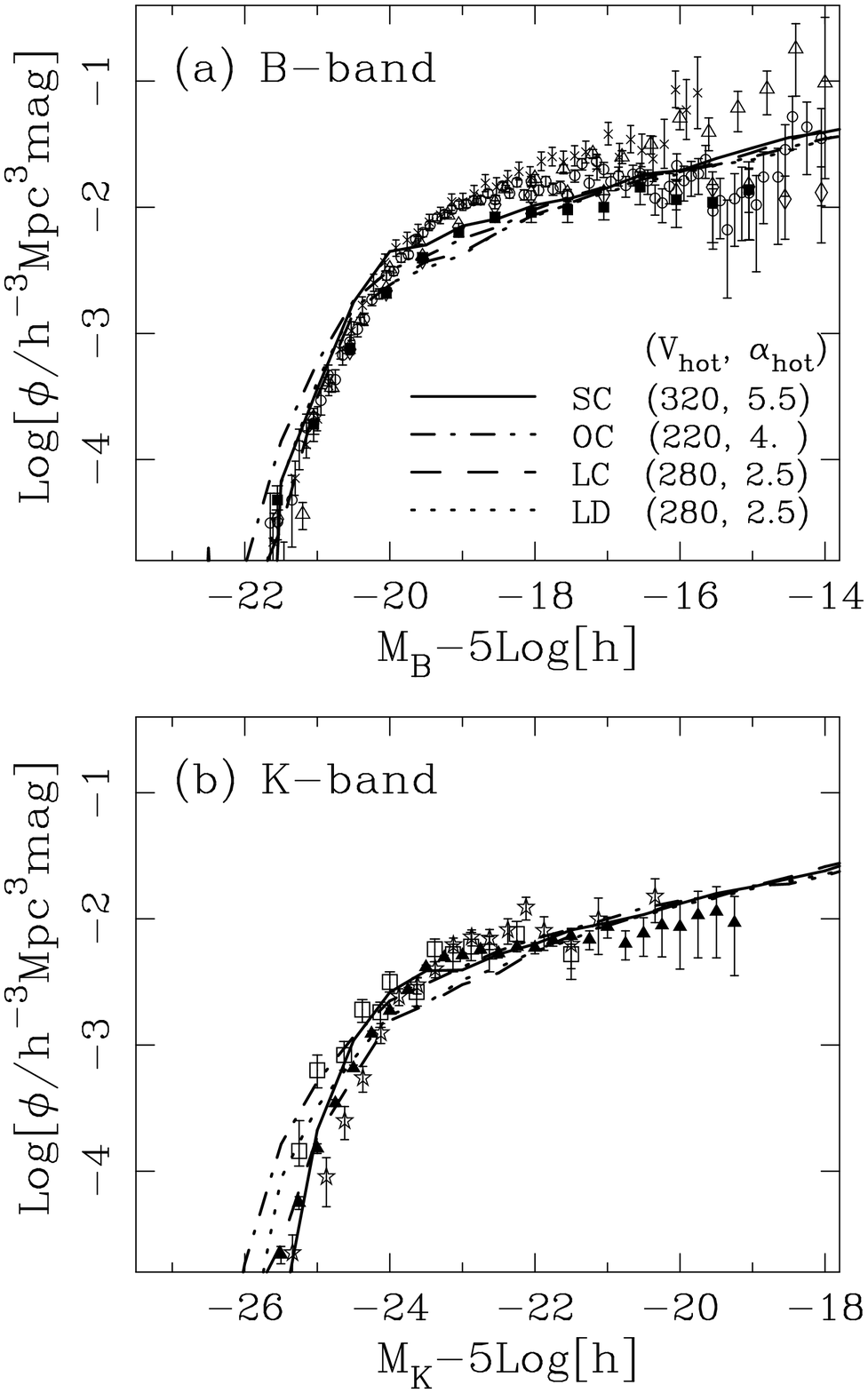}
\caption{Local luminosity functions in the (a) $B$ band and (b) $K$
 band.  The solid, dot-dashed, dashed, and dotted lines indicate the SC,
 OC, LC and LD models, respectively.  Symbols with errorbars in (a)
 indicate the observational data from APM (Loveday et al. 1992, {\it
 filled squares}), ESP (Zucca et al. 1997, {\it open triangles}),
 Durham/UKST (Ratcliffe et al. 1998, {\it open diamonds}), 2dF (Folkes
 et al. 1999, {\it open circles}), and SDSS (Blanton et al. 2000, {\it
 crosses}).  Note that the SDSS luminosity function shown here is that
 with the same detection limit as employed in the 2dF survey.  Symbols
 in (b) indicate the data from Mobasher et al.  (1993, {\it open
 squares}), Gardner et al. (1997, {\it open stars}), and 2MASS (Cole et
 al. 2000b, {\it filled triangles}).  }
\label{fig:lf}
\end{inlinefigure}

\begin{inlinefigure}
\includegraphics[width=8cm]{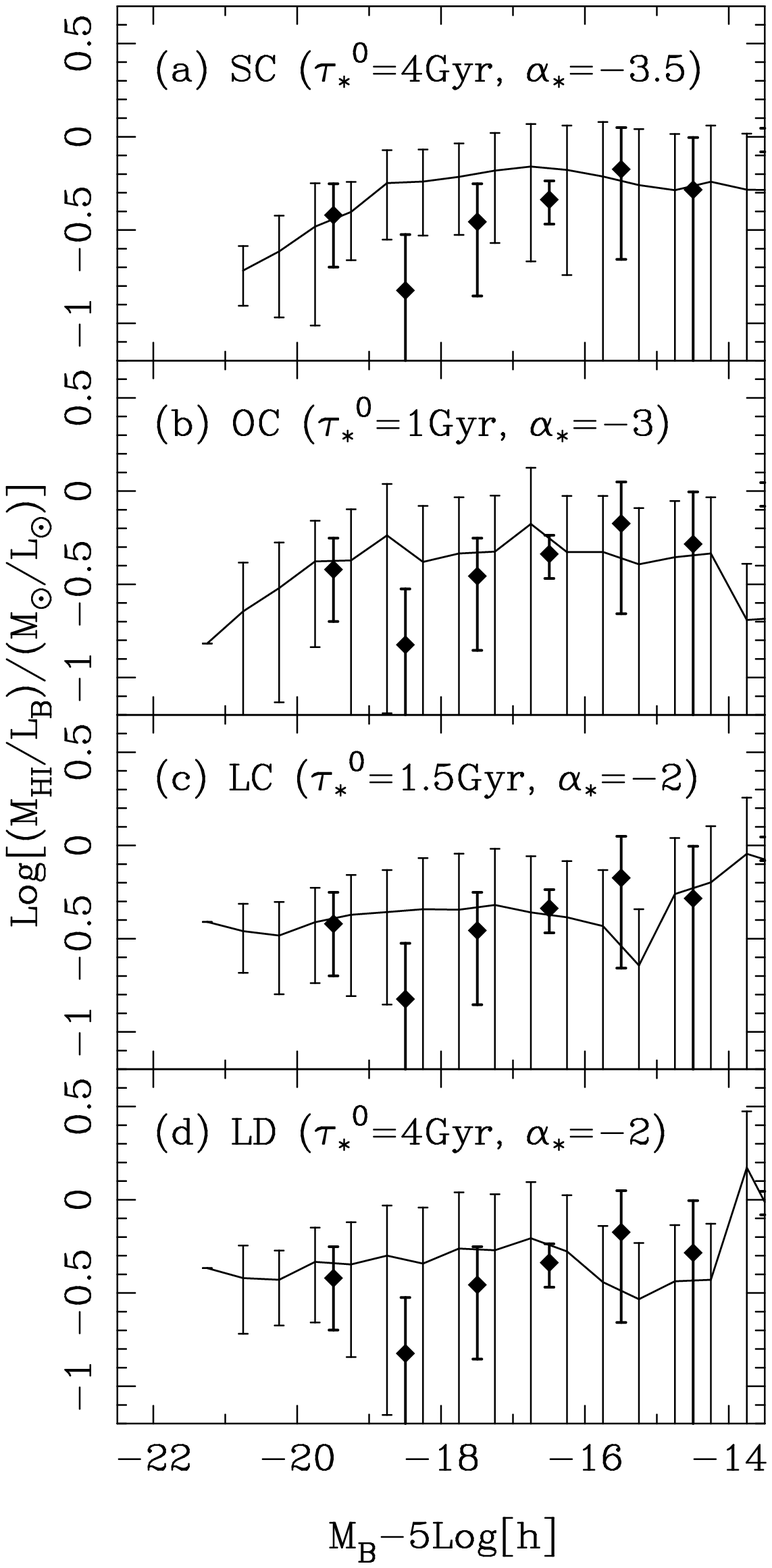}
\caption{Cold gas mass relative to $B$-band luminosity of spiral
 galaxies.  Solid curves in the four panels from top to bottom show the
 SC, OC, LC and LD models, respectively.  Errorbars denote 1$\sigma$
 scatters.  Filled diamonds indicate the observational data for atomic
 neutral hydrogen taken from Huchtmeier \& Richter (1988).  In the
 models, the cold gas consists of all species of elements, therefore its
 mass is multiplied by 0.75, i.e., $M_{\rm HI}=0.75M_{\rm cold}$, which
 corresponds to the fraction of hydrogen.  Because the observational
 data denote only atomic hydrogen, they should be interpreted as lower
 limits of the ratio.  }
 \label{fig:gas}
\end{inlinefigure}

\onecolumn
\begin{inlinefigure}
\includegraphics[width=8cm]{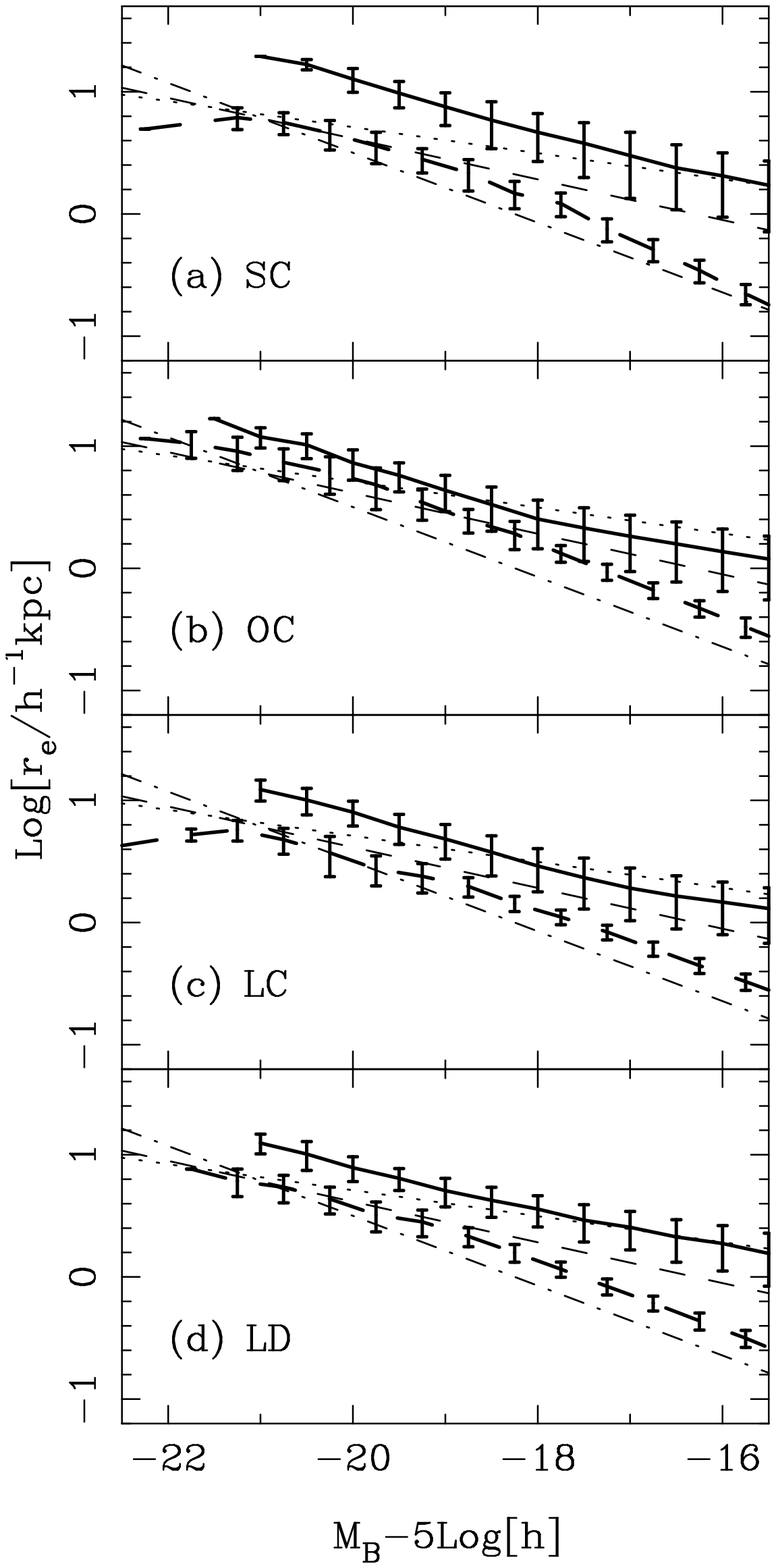}
\caption{Galaxy sizes.  The thick solid lines and the thick dashed lines show
theoretical results for effective radii of late-type and early-type galaxies,
respectively. Errorbars denote 1$\sigma$ scatters.  The thin dotted lines,
the thin dashed lines, and the thin dot-dashed lines show the observed
best-fit relations given by TY00 for spirals, dwarf
ellipticals and compact ellipticals, respectively.  The original data
are given by Impey et al. (1996) for spirals and Bender et al. (1992)
for ellipticals.  } \label{fig:rad}
\end{inlinefigure}

\begin{inlinefigure}
\includegraphics[width=8cm]{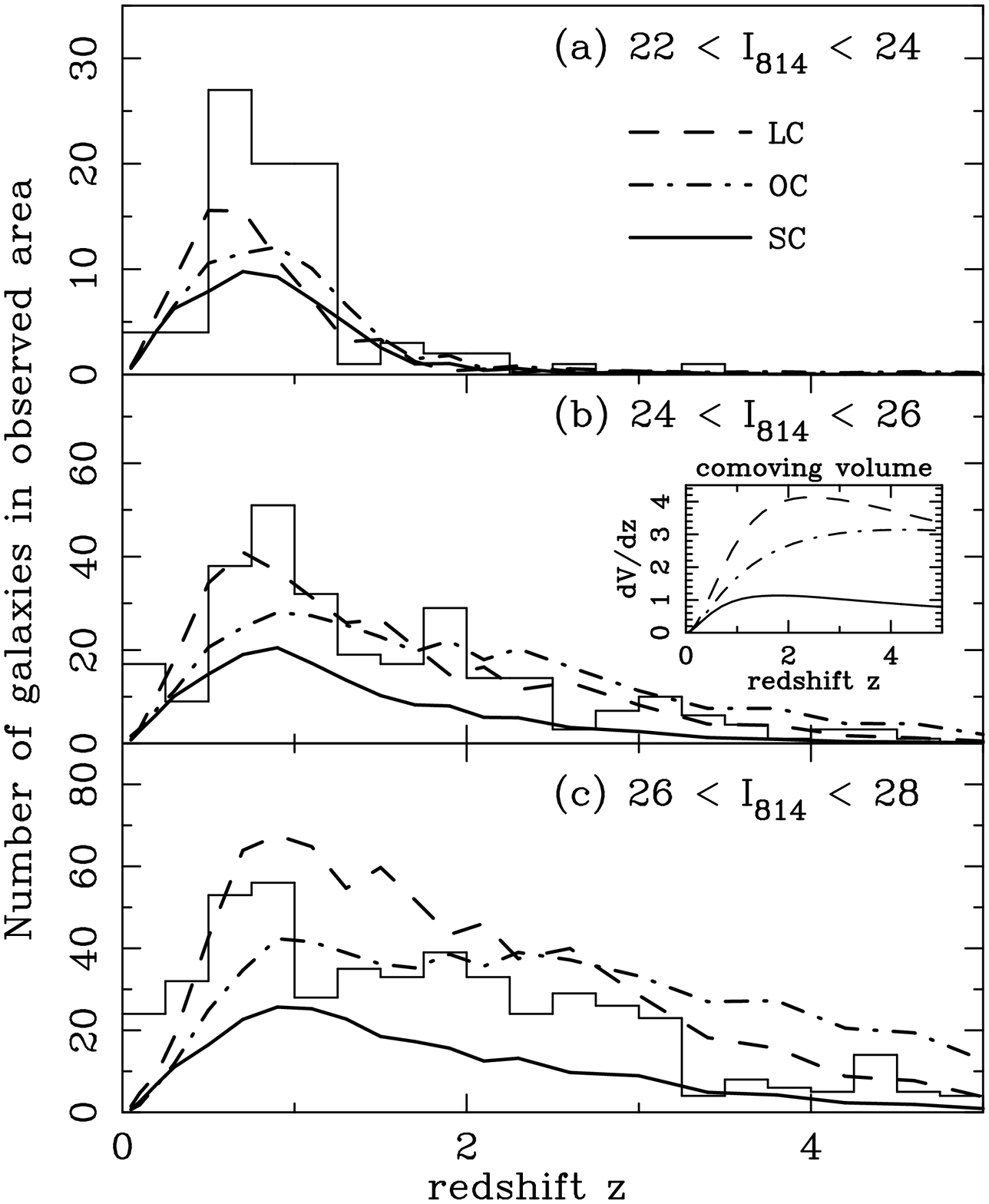}
\caption{Redshift distributions for various cosmological models.  The
 types of lines are the same as Fig. \ref{fig:counts}, but for only
 models including the effects of the selection and the absorption by the
 \ion{H}{1} clouds.  The histograms indicate the photometric redshift
 distribution of the HDF.  For reference, the redshift dependence of the
 comoving volume element $dV/dz$ in units of $10^{6}h^{-3}$Mpc$^{3}$ is
 shown in the middle panel for the three cosmological models.  The types
 of lines are the same as corresponding cosmological models for the
 redshift distributions of galaxies.}
\label{fig:zdist}
\end{inlinefigure}

\begin{inlinefigure}
\includegraphics[width=8cm]{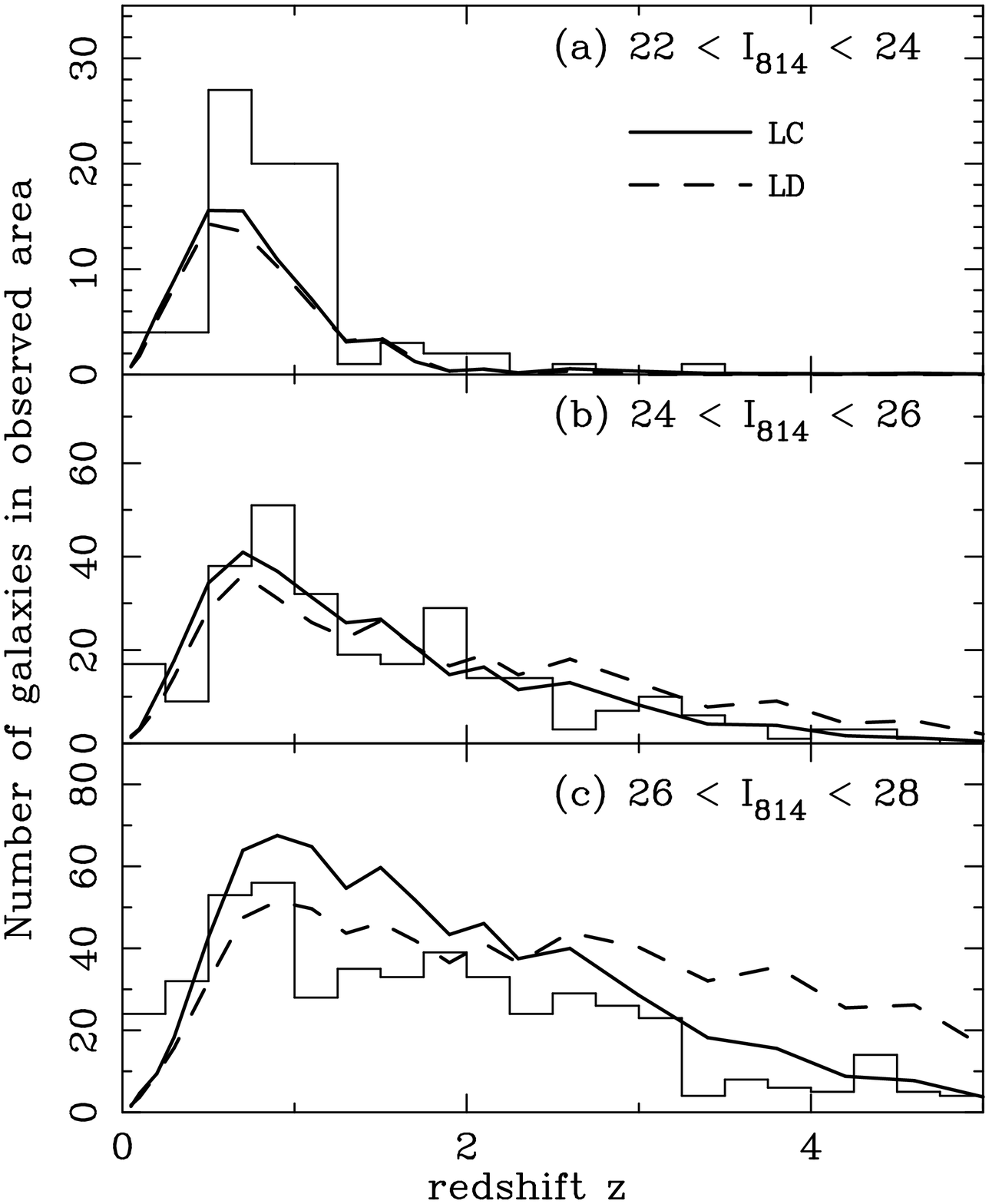}
\caption{Redshift distributions.  The solid lines and the dashed
lines indicate the LC and LD models, respectively.  The histograms 
indicate the photometric redshift distribution of the HDF. }
\label{fig:z_sft}
\end{inlinefigure}

\begin{inlinefigure}
\includegraphics[width=8cm]{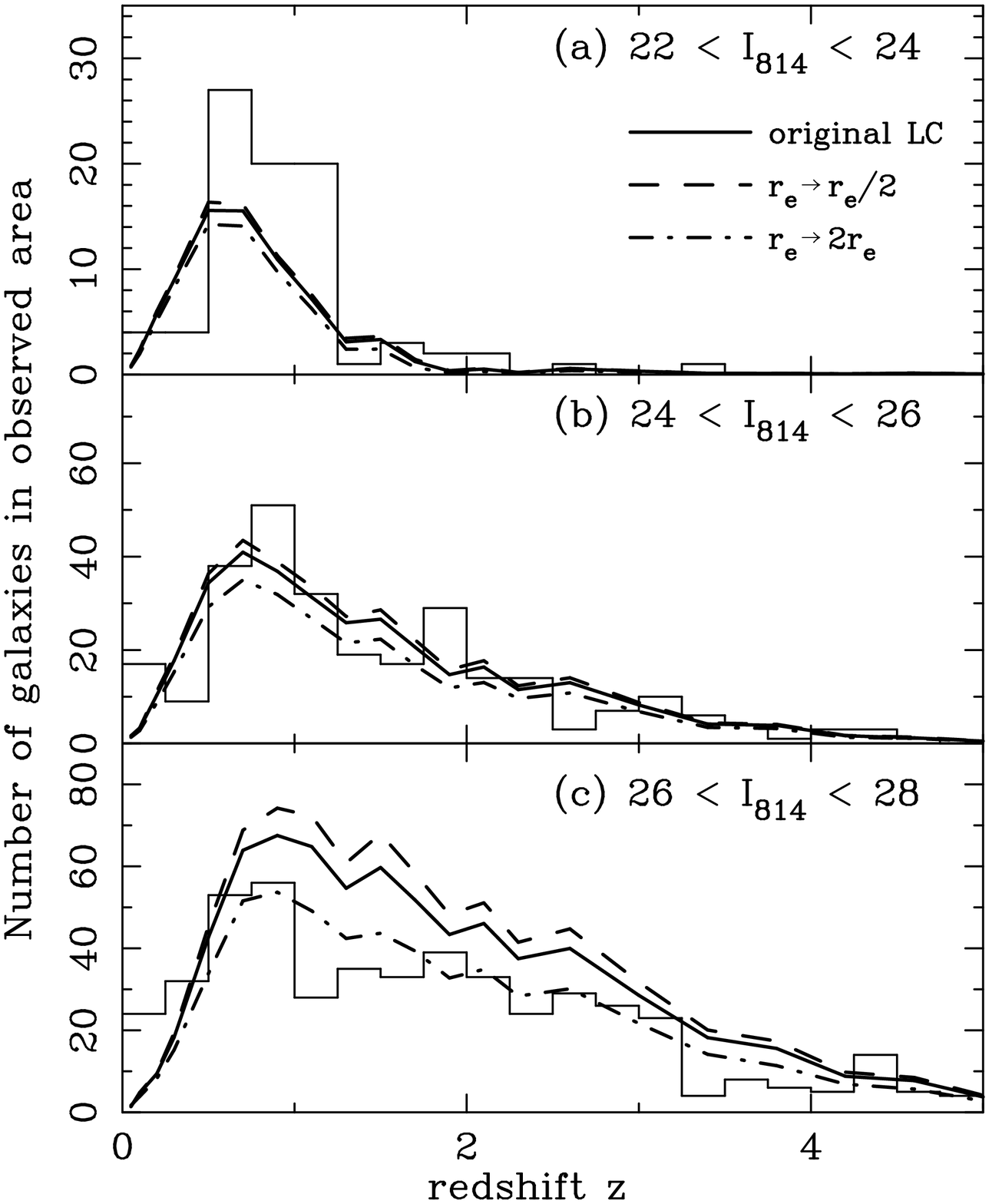}
\caption{Redshift distributions.  The thick solid lines indicate the LC
model.  The dashed and dot-dashed lines denote the high and low surface
brightness models for the model LC, respectively.  The histograms
indicate the photometric redshift distribution of the HDF. }
\label{fig:z_rad_lcdm}
\end{inlinefigure}

\begin{inlinefigure}
\includegraphics[width=8cm]{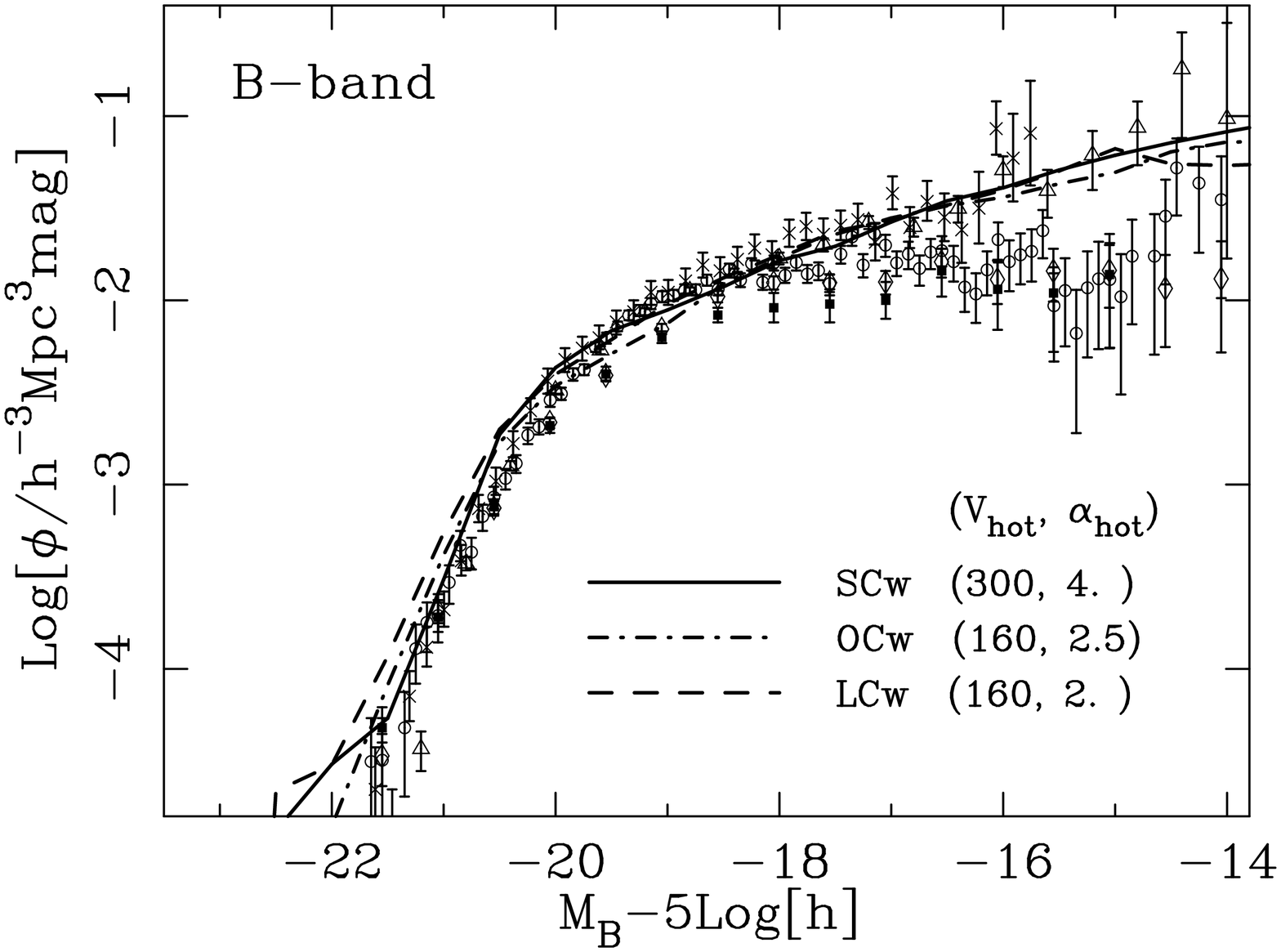}
\caption{Local luminosity function in the $B$-band.  The solid,
dot-dashed and dashed lines indicate the SCw, OCw and LCw models,
respectively.  The types of symbols denote observations and are the same
as Fig. \ref{fig:lf}a.}
 \label{fig:lf_fb}
\end{inlinefigure}

\begin{inlinefigure}
\includegraphics[width=8cm]{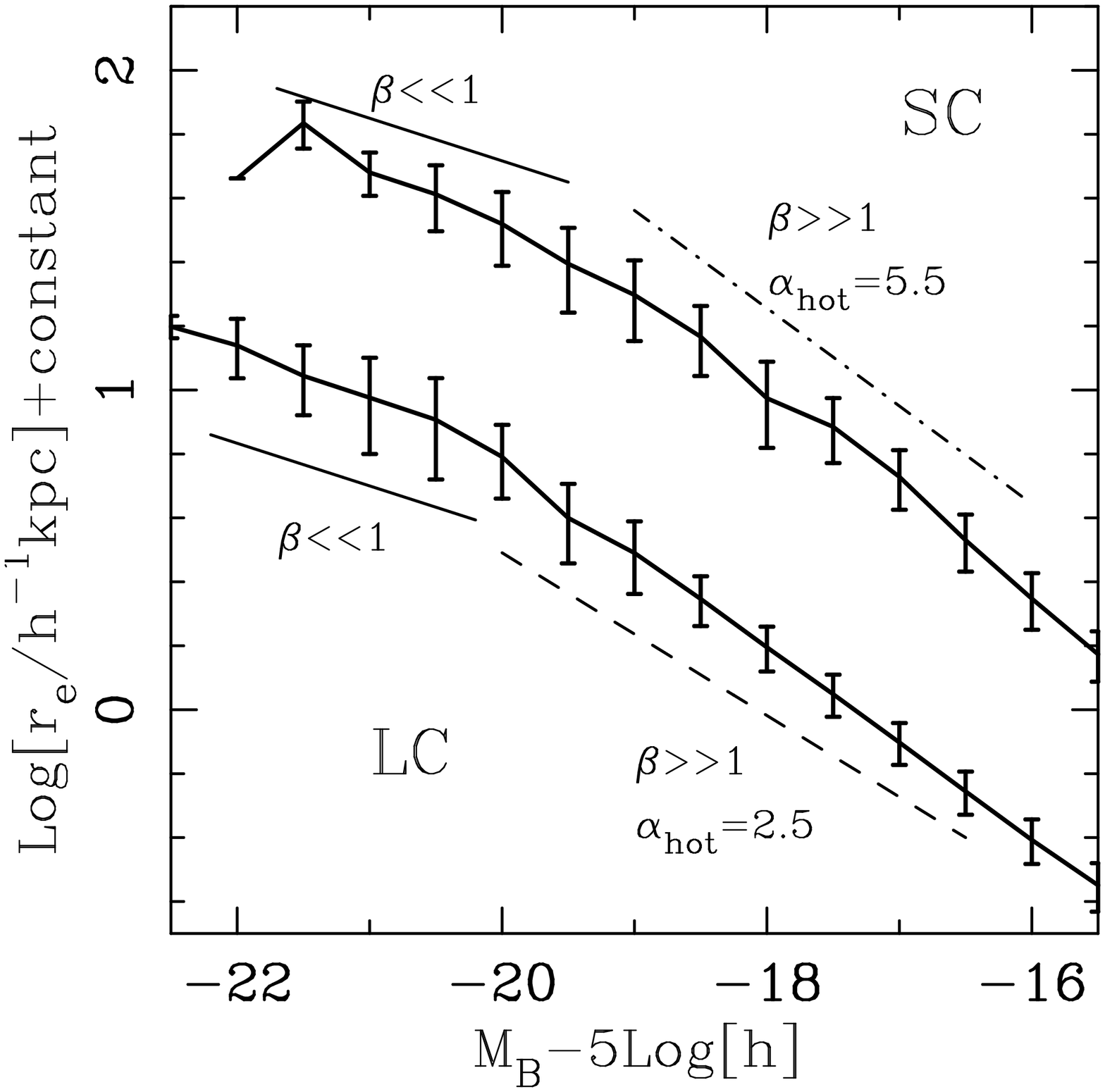}
\caption{Galaxy sizes in arbitrary unit.  The upper and lower thick
lines indicate effective radius of early-type galaxies in the SC and LC
models, respectively.  The thin solid lines indicate the case of
$\beta\ll 1$ and the thin dot-dashed line and the thin dashed line
indicate the case of $\beta\gg 1$ with $\alpha_{\rm hot}=5.5$ and 2.5,
respectively.}
\label{fig:rad_ana}
\end{inlinefigure}

\begin{inlinetable}
\begin{center}  
\caption{Model Parameters}  
\label{tab:astro}
\begin{tabular}{ccccccccccccc}
\hline
\hline
& \multicolumn{4}{c}{cosmological parameters} &
& \multicolumn{7}{c}{astrophysical parameters} \\
\cline{2-5} \cline{7-13}
CDM Model & $\Omega_{0}$ & $\Omega_{\Lambda}$ &$h$ &$\sigma_{8}$ & &
$V_{\rm hot}$ (km~s$^{-1}$)
& $\alpha_{\rm hot}$ & $\tau_{*}^{0}$ (Gyr)
& $\alpha_*$ & $f_{\rm bulge}$ & $f_{\rm b}$ & $\Upsilon$\\
\hline
SC & 1 & 0 &0.5&0.67&& 320 & 5.5 & 4   & -3.5 & 0.2 & 1   & 1.\\
OC &0.3& 0 &0.6& 1  && 220 & 4   & 1   & -3   & 0.5 & 0.5 & 1.5\\
LC &0.3&0.7&0.7& 1  && 280 & 2.5 & 1.5 & -2   & 0.5 & 0.5 & 1.5\\
LD &0.3&0.7&0.7& 1  && 280 & 2.5 & 4   & -2   & 0.5 & 0.5 & 1.5\\
\hline
\end{tabular}
\end{center}
\end{inlinetable}
   
\begin{inlinetable}
\begin{center}  
\caption{Parameters for Weak Feedback Models}  
\label{tab:astro2}
\begin{tabular}{ccccccccccccc}
\hline
\hline
& \multicolumn{4}{c}{cosmological parameters} &
& \multicolumn{7}{c}{astrophysical parameters} \\
\cline{2-5} \cline{7-13}
CDM Model & $\Omega_{0}$ & $\Omega_{\Lambda}$ &$h$ &$\sigma_{8}$ & &
$V_{\rm hot}$ (km~s$^{-1}$)
& $\alpha_{\rm hot}$ & $\tau_{*}^{0}$ (Gyr)
& $\alpha_*$ & $f_{\rm bulge}$ & $f_{\rm b}$ & $\Upsilon$\\
\hline
SCw & 1 & 0 &0.5&0.67&&  300 & 4   & 4   & -2.5 & 0.2 & 1   & 1\\
OCw &0.3& 0 &0.6& 1  &&  160 & 2.5 & 1.5 & -1.5 & 0.5 & 0.5 & 1.8\\
LCw &0.3&0.7&0.7& 1  &&  160 & 2   & 1.5 & -1.5 & 0.5 & 0.5 & 1.5\\
\hline
\end{tabular}
\end{center}
\end{inlinetable}

\end{document}